\documentclass[journal]{IEEEtran}
\usepackage{times}  
\usepackage{helvet}  
\usepackage{courier}  
\usepackage[hyphens]{url}  
\usepackage{graphicx} 
\usepackage{amsfonts}
\usepackage{amssymb}
\usepackage{amsthm}
\usepackage{amsmath}
\usepackage{subcaption}
\usepackage{siunitx}
\usepackage{xcolor}
\usepackage{subfiles} 
\usepackage{cite}
\usepackage{tabularx}
\usepackage[
    colorlinks=false,   
    pdfborder={0 0 0}    
]{hyperref}
\usepackage{threeparttable}
\usepackage{booktabs}
\usepackage[table]{xcolor}
\usepackage{colortbl}
\usepackage{array, makecell}
\usepackage{multirow}
\usepackage[noend]{algpseudocode}
\theoremstyle{bolddefinition}

\ifCLASSINFOpdf
\else
\fi
\begin{document}
%
\title{LA-RL: Language Action-guided Reinforcement Learning with Safety Guarantees for Autonomous Highway Driving}
%
%
%

\author{Yiming Shu, Jiahui Xu, Jiwei Tang, Ruiyang Gao, Chen Sun
\thanks{This work was supported by the Research Grants Council of Hong Kong under Grant No. 27206525.}
\thanks{Y. Shu, J. Xu, J. Tang, R. Gao and C. Sun are with the Department of Data and Systems Engineering, the University of Hong Kong, Hong Kong SAR. (Email: {\texttt{c87sun@hku.hk}}) }}
%
%

\markboth{IEEE Transactions on Vehicular Technology}%
{Shell \MakeLowercase{\textit{et al.}}: Bare Demo of IEEEtran.cls for IEEE Journals}
%



\maketitle

\begin{abstract}
Autonomous highway driving demands a critical balance between proactive, efficiency-seeking behavior and robust safety guarantees. This paper proposes Language Action-guided Reinforcement Learning (LA-RL) with Safety Guarantees, a novel framework that integrates the semantic reasoning of large language models (LLMs) into the actor-critic architecture with an improved safety layer. Within this framework, task-specific reward shaping harmonizes the dual objectives of maximizing driving efficiency and ensuring safety, guiding decision-making based on both environmental insights and clearly defined goals. To enhance safety, LA-RL incorporates a safety-critical planner that combines model predictive control (MPC) with discrete control barrier functions (DCBFs). This layer formally constrains the LLM-informed policy to a safe action set, employs a slack mechanism that enhances solution feasibility, prevents overly conservative behavior and allows for greater policy exploration without compromising safety. Extensive experiments demonstrate that it significantly outperforms several current state-of-the-art methods, offering a more adaptive, reliable, and robust solution for autonomous highway driving. Compared to existing SOTA, it achieves approximately 20$\%$ higher success rate than the knowledge graph (KG) based baseline and about 30$\%$ higher than the retrieval augmented generation (RAG) based baseline. In low-density environments, LA-RL achieves a 100$\%$ success rate. These results confirm its enhanced exploration of the state-action space and its ability to autonomously adopt more efficient, proactive strategies in complex, mixed-traffic highway environments.
\end{abstract}


Videos of our experiments: \url{https://github.com/YimingShu-teay/LA-RL}

%
\IEEEpeerreviewmaketitle

\begin{figure*}[t]
  \centering
  \includegraphics[scale=0.275,trim=0 -2 0 0, clip]{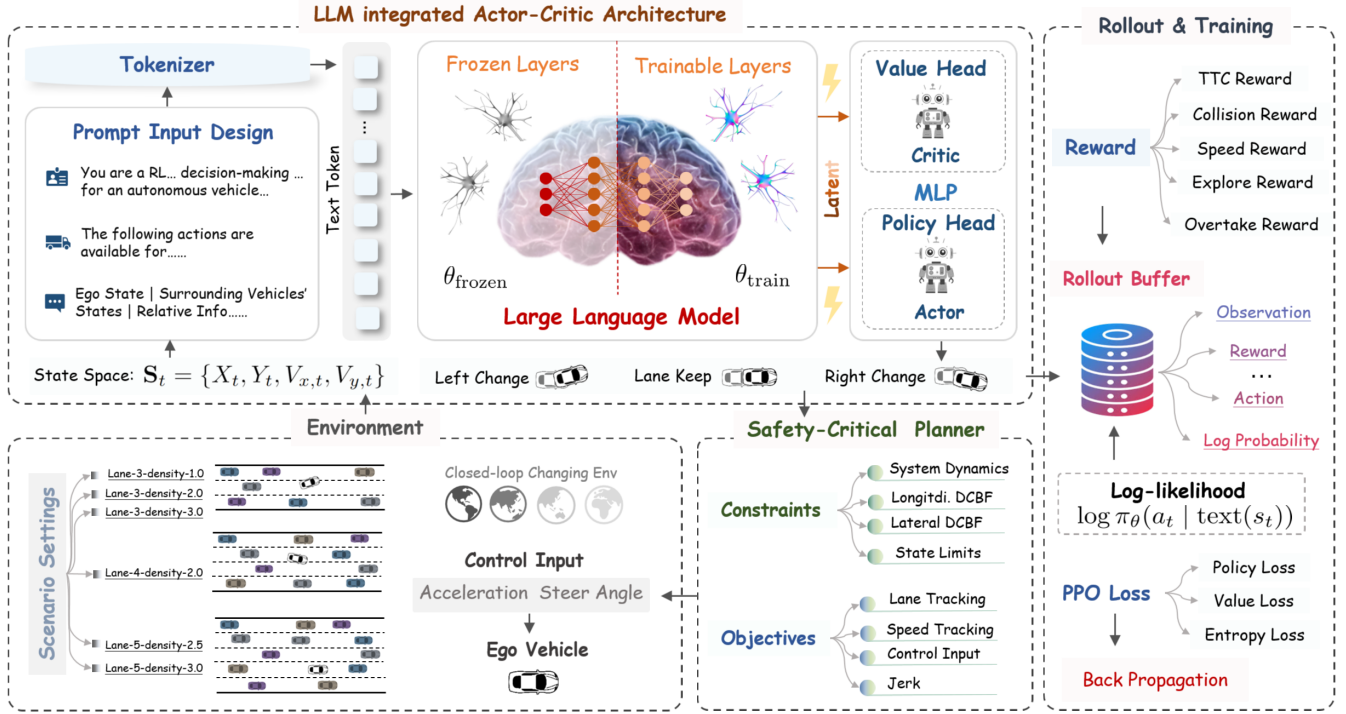}
  \caption{The overall framework of LA-RL. The framework combines LLM with RL, where the LLM serves as the language-informed decision-maker, processing language-guided information. The safety-critical planner integrates Model Predictive Control (MPC) and Discrete Control Barrier Functions (DCBF) to generate optimized control inputs, ensuring safe and robust operation with slack mechanism. Multiple reward components, including TTC, collision, speed, exploration, and overtaking rewards, are utilized to guide the autonomous vehicle's decision-making in various complex settings.}
  \label{Outline}
\end{figure*}

\section{Introduction}
The development of autonomous driving has progressed from rule-based~\cite{shu2025agile}, modular systems to data-driven, end-to-end (e2e) architectures driven by the goal towards unified optimization and scaling in mixed-autonomy traffic. 
Yet, the data-driven e2e strategies introduce new challenges such as bias and overfitting~\cite{wen2024dilu}, tendency toward simplistic behavioral imitation without genuine environmental understanding~\cite{zhou2025autovla} and lack of robust, verifiable safety guarantees. It is crucial to explore a new planning paradigm which treats safety as a foundational, non-negotiable constraint while transcending mere imitation by enabling vehicles to actively interpret complex physical scenarios and reason about efficient, proactive strategies during environmental exploration.

Large language models boost scene understanding through semantic interpretation and contextual reasoning~\cite{zhang2025lead}. Its knowledge-driven capability in scenario understanding and decision-making has attracted substantial attention and interest in its integration in autonomous driving systems~\cite{brown2020language,gyevnar2024building,wei2024editable}. Although LLMs bring along many advantages, without expert guidance, limitations in reliability may culminate in suboptimal decision-making performance and low reliability for real-world scenarios~\cite{wang2025hybrid}.

Autonomous driving systems industry utilizes deep reinforcement learning (DRL) planning since its good performance solving Markov Decision Process (MDP) with scenario encoding, which presents salient efficacy in autonomous driving decision-making~\cite{pang2024large,jia2024learning}. However, the intricate extraction of features for DRL can introduce subjective biases~\cite{xiong2025flag}, which may impart DRL with traits such as poor scenario comprehension. 

The integration of DRL and LLM has led to a surge of recent works~\cite{xu2025tell,chen2025hcrmp,sun2024optimizing,chen2024driving}. For instance, RAPID~\cite{wu2024robust} integrates LLM knowledge distillation and adaptive RL for robust autonomous driving.  A ``fast-slow'' architecture is proposed, combining LLM for high-level instruction parsing and RL for real-time low-level decision-making~\cite {xu2025towards}. Nevertheless, ensuring safety is still a precarious problem of the DRL-based framework, as it is inadequate to mitigate collision risks through interaction during exploration~\cite {shu2023safety,shu2025agile}.

To address the issues above, we propose a monolithic framework named \textbf{L}anguage \textbf{A}ction-guided \textbf{R}einforcement \textbf{L}earning with Safety Guarantees (LA-RL), as illustrated in Fig.~\ref{Outline}.  LA-RL employs environmental condition description as the input, leveraging the strengths of LLM to interpret the environment deeply. Inspired by~\cite{xiong2025flag}, unlike conventional Proximal Policy Optimization (PPO) implementations, we use the LLM as the backbone within the actor-critic architecture, enabling language-informed decision-making. Considering the AV as an agent, the combination of LLM and DRL offers an innovative perspective, providing both knowledge and explicit task-aligned reward guidance. Once the action is determined, the safety-critical planner, consisting of model predictive control (MPC) and discrete control barrier functions (DCBFs), will generate control inputs through a finite-horizon convex optimization problem. Extensive experiments and evaluations highlight LA-RL's ability to prioritize safety while exploring advantages in spatial and speed capabilities.
Overall, the main contributions of our work are as follows:
\begin{itemize}
\item We propose LA-RL, a framework that leverages an LLM as a policy generator with partially unfrozen layers for online PPO training. This approach enables the model to adaptively learn traffic dynamics while retaining prior knowledge, guided by a novel reward function designed for highway driving.
\item We integrate a safety-critical MPC-DCBF planner and a slack mechanism to enhance both safety and feasibility. This encourages greater exploration by preventing the model from overly conservative avoidance, thus enabling it to learn sagacious driving decisions.
\item In rigorous evaluation across diverse multi-lane, high-density highways, our LA-RL framework significantly outperforms SOTA. It achieves a 20$\%$ and 30$\%$ higher success rate than the KG-based baseline and the RAG-based baseline, respectively. Furthermore, LA-RL attains a 100$\%$ success rate with greater driving progress in low-density scenarios, demonstrating robust performance across all conditions.
\end{itemize}

The rest of this paper is organized as follows. Section~\ref{sec:related_works} provides a review of the relevant literature. Section~\ref{sec:Preliminaries} presents the preliminary of this paper. Section~\ref{sec:Methodology} introduces the LA-RL framework. Section~\ref{sec:experiment} discusses the experimental results of the proposed method. Section~\ref{sec:conclusion} concludes the paper.

\section{Related Works}
\label{sec:related_works}
\subsubsection{DRL-based Planning Systems}
As a powerful approach, DRL has been applied to numerous scenarios in autonomous driving, spanning from adaptive cruise control (ACC) to multi-agent driving coordination tasks~\cite{hua2025multi,wang2024multiagent}. ~\cite{chen2022dqn} presents ES-DQN, a deep Q-learning method that improves vehicle speed control in uncertain cut-in scenarios. Originally proposed by OpenAI~\cite{schulman2017proximal}, Proximal Policy Optimization (PPO) has become one of the most widely used algorithms in RL. To address the challenges of safe lane changing,~\cite{ye2020automated} presents a safety-aware PPO method with an aborting action. Pad-AI~\cite{jia2024learning} introduces a perception-aware RL framework for occluded scenarios, combining vectorized observations, semantic motion primitives, and safe interaction. However, the above methods hardly consider the environmental comprehension of the model.

\subsubsection{Interpretable LLM-integrated Planning Methods}
LLM helps AVs to interpret traffic conditions better. DiLu~\cite{wen2024dilu} introduces a knowledge-driven framework that enables LLM-based agents to reason, reflect, and improve through accumulated driving experiences. LC-LLM~\cite{peng2025lc} leverages LLMs and chain-of-thought (CoT) reasoning to produce interpretable lane change intention and trajectory predictions by framing the task as language modeling.  ~\cite{lan2024traj} proposes Traj-LLM, a prompt-free LLM-based framework with lane-aware learning and multi-modal decoding for trajectory prediction. However, some of them may suffer from issues such as hallucinations or a lack of behavioral guidance.

\subsubsection{Safety-critical Planning} 
Safety-critical planning plays an important role in autonomous driving. CBFs and MPC are two widely recognized approaches to achieve safety-critical planning and control.
A safety-critical adaptive cruise control (ACC) system was initially proposed by \cite{ames2014control} with one-step control, and was later extended by \cite{he2021rule} for rule-based lane-changing strategy.  Instead of controlling for a single step, MPC optimizes over a receding horizon with explicit hard constraints. \cite{zeng2021safety} first introduced the integration of MPC with discrete-time control barrier functions (MPC-DCBF), highlighting the tradeoff between safety and feasibility. Further applications of MPC-DCBF include \cite{shu2023safety,he2022autonomous,zhang2024courteous}, which demonstrated its effectiveness in multiple driving scenarios. However, many of these works struggle with the nonlinear programming problem (NLP) and the solution feasibility.

\section{Preliminaries}
\label{sec:Preliminaries}
In this section, we provide the theoretical background for reinforcement learning (RL) and proximal policy optimization (PPO) in Section~\ref{sec:pre_RL_PPO}, model predictive control (MPC) in Section~\ref{sec:MPC}, and control barrier functions (CBF) in Section~\ref{sec:CBF}.

\subsection{Reinforcement Learning and Proximal Policy Optimization}
\label{sec:pre_RL_PPO}
The Markov Decision Process (MDP) is formally defined as a tuple $\mathcal{M} = (\mathcal{S}, \mathcal{A}, \mathcal{T}, \mathcal{R}, \gamma)$, where $\mathcal{S}$ and $\mathcal{A}$ denote the state and action spaces, respectively. The dynamics of the environment are captured by two core functions: the transition function $\mathcal{T}(s'|s, a): S \times \mathcal{A} \times \mathcal{S} \rightarrow [0, 1]$, which defines the probability of transitioning to state $s'$ from state $s$ after taking action $a$, and the reward function $\mathcal{R}(s, a, s'): S \times \mathcal{A} \times S \rightarrow \mathbb{R}$, which specifies the immediate reward received for this transition. $\gamma \in (0,1)$ refers to the discount factor for future reward. To find the optimal policy, the state-value and state-action value functions are defined as follows:
\begin{equation}
V^{\pi}(s_t) = \mathbb{E}_{a_t \sim \pi(\cdot|s_t)} [Q^{\pi}(s_t, a_t)],
\end{equation}
\begin{equation}
Q^{\pi}(s_t, a_t) = r_t + \gamma \mathbb{E}_{s_{t+1} \sim P(\cdot|s_t, a_t)} [V^{\pi}(s_{t+1})],
\end{equation}
where $\pi(a_t|s_t)$ is the policy. The optimal policy satisfies
$\pi^* = \arg \max_{\pi} V^{\pi}(s_t)$. DRL addresses the challenge of accurately estimating value functions to obtain $\pi^*$ by using neural networks to approximate both the policy and value functions.

This paper employs PPO, which updates the actor using two concurrent networks to maintain old and new policies, driven by the policy ratio $r_t(\theta) = \frac{\pi_\theta(a_t|s_t)}{\pi_{\theta_{\mathrm{old}}}(a_t|s_t)}$. PPO introduces a new clip mechanism, which can effectively reduce the number of computation steps while limiting the magnitude of policy update, and it is defined as follows:
\begin{equation}
\label{eq:ppo_clip_loss}
\begin{aligned}
L_t^{CLIP}(\theta) &= \mathbb{E}_t[\min(r_t(\theta) \hat{A}_t, \\
&\quad \mathrm{clip}(r_t(\theta), 1 - \epsilon, 1 + \epsilon) \hat{A}_t)].
\end{aligned}
\end{equation}
The purpose of setting $1 - \epsilon, 1 + \epsilon$ is to specify the magnitude of the policy update, $\epsilon$ is a hyperparameter, Generalized Advantage Estimation (GAE) is utilized to compute the advantage function $\hat{A}_t = \delta_t + (\gamma \lambda) \cdot \hat{A}_{t+1}$, $\delta_t$ is the TD error $r_t + \gamma V(s_{t+1}) - V(s_t)$. The policy-based approach parameterizes the policy directly and updates the parameters of the policy through the policy gradient to maximize rewards. The PPO loss is defined as follows:
\begin{equation}
\begin{split}
L_t^{CLIP+VF+S}(\theta) &= \mathbb{E}_t \left[ L_t^{CLIP}(\theta) \right] \\
 &- c_1 \mathbb{E}_t \left[ L_t^{VF}(\theta) \right] + c_2 \mathbb{E}_t \left[ S[\pi_{\theta}](s_t) \right].
\end{split}
\label{PPO_loss}
\end{equation}
where $\theta$ is the policy parameter. $L_t^{CLIP}$ is the clipped loss. $c_1$, $c_2$ are coefficients. $S$ denotes an entropy bonus. $L_t^{VF}$ is a squared-error loss $(V_{\theta}(s_t) - V_t^{targ})^2$. The PPO penalty variant employs a clipped surrogate objective, utilizes a learned state-value function $V(s)$, and incorporates an entropy bonus to encourage exploration.

\subsection{Model Predictive Control}
\label{sec:MPC}
We consider the problem of regulating the discrete-time control system in the form:
\begin{equation}
\mathbf{x}_{t+1} = f(\mathbf{x}_t, \mathbf{u}_t),
\label{pre:system}
\end{equation}
where $\mathbf{x}_t \in \mathcal{X} \subset \mathbb{R}^n$ represents the state of the system at time step $t \in \mathbb{Z}^+$, $\mathbf{u}_t \in \mathcal{U} \subset \mathbb{R}^m$ is the control input, and $f$ is locally Lipschitz.
Given the state $\mathbf{x}_t$ at time $t$, a finite-time optimal control problem is solved. Obstacle avoidance constraints are incorporated as distance constraints in the following formulation:
\begin{equation}
\begin{split}
 & \underset{\mathbf{u}_t:t+N-1|t,\mathbf{x}_t:t+N|t}{\arg\min}  p(\mathbf{x}_{t+N|t}) + \sum_{k=0}^{N-1} q(\mathbf{x}_{t+k|t}, \mathbf{u}_{t+k|t})  \\
\text{s.t.} \quad & \mathbf{x}_{t+k+1|t} = f(\mathbf{x}_{t+k|t}, \mathbf{u}_{t+k|t}), \quad k = 0, \dots, N-1 \\
& \mathbf{x}_{t+k|t} \in \mathcal{X}, \mathbf{u}_{t+k|t} \in \mathcal{U}, \quad k = 0, \dots, N-1 \\
& \mathbf{x}_{t|t} = \mathbf{x}_t, \\
& h(\mathbf{x}_{t+k|t}) \geq 0, \quad k = 0, \dots, N-1. 
\end{split}
\raisetag{4em} 
\label{mpc}
\end{equation}
MPC iteratively solves a finite-horizon optimal control problem. At each time step $t$, the optimal control sequence $\mathbf{u}_{t:t+N-1|t}^*$ is computed by minimizing a cost function comprising stage cost $q(\mathbf{x}_{t+k|t}, \mathbf{u}_{t+k|t})$ and terminal cost $p(\mathbf{x}_{t+N|t})$ subject to constraints. Only the first control action $\mathbf{u}(t) = \mathbf{u}_{t|t}^*(\mathbf{x}_t)$ is applied to Eq.~\eqref{pre:system}. 

\subsection{Control Barrier Functions}
\label{sec:CBF}


Within the safety-critical control paradigm, the safe set $\mathcal{C}$ is given by the superlevel set of a continuously differentiable function $h : \mathcal{X} \subset \mathbb{R}^n \to \mathbb{R}$:
\begin{equation}
\mathcal{C} = \{\mathbf{x} \in \mathcal{X} \subset \mathbb{R}^n : h(\mathbf{x}) \geq 0\}.
\end{equation}
We refer to $\mathcal{C}$ as a safe set. The function $h$ is a control barrier function (CBF)~\cite{ames2014control} if $\frac{\partial h}{\partial \mathbf{x}} \neq 0$ for all $\mathbf{x} \in \partial \mathcal{C}$ and there exists an extended class $\mathcal{K}_\infty$ function $\gamma$ such that for the control system (1), $h$ satisfies
\begin{equation}
\exists \mathbf{u} \text{ s.t. } \dot{h}(\mathbf{x}, \mathbf{u}) \geq -\gamma(h(\mathbf{x})), \ \gamma \in \mathcal{K}_\infty.
\end{equation}
The extension of this condition to the discrete-time domain takes the following form:
\begin{equation}
\Delta h(\mathbf{x}_k, \mathbf{u}_k) \geq -\gamma h(\mathbf{x}_k), \quad 0 < \gamma \leq 1,
\label{pre:dcbf}
\end{equation}
where $\Delta h(\mathbf{x}_k, \mathbf{u}_k) := h(\mathbf{x}_{k+1}) - h(\mathbf{x}_k)$. The constraint in \eqref{pre:dcbf} implies that $h(\mathbf{x}_{k+1}) \geq (1 - \gamma)h(\mathbf{x}_k)$, meaning that the lower bound of the control barrier function $h(\mathbf{x})$ decreases exponentially at a rate of $1 - \gamma$.

\section{Methodology}
\label{sec:Methodology}
In this section, we will first introduce the reward design in Section \ref{subsec:reward_design}. Next, the LLM integrated actor-critic architecture will be presented in Section \ref{subsec:actor-critic}. Finally, we will introduce the safety-critical planner in Section \ref{subsec:safety-critical}.

\subsection{Reward Design}
\label{subsec:reward_design}
Exquisite rewards shape targets for a decision's balance of efficiency and safety.  Many existing works utilize the speed difference between a desired velocity and a collision as the efficiency reward and safety reward, respectively, which may cause erratically negative exploration until the collision occurs, or result in overly conservative behaviors for collision avoidance. Instead of the common design strategies, we devised three partitions for the efficiency reward and two parts for the safety reward:
\begin{equation}
    R = \delta_{scal}(R_{\text{efficiency}} + R_{\text{safety}})*\delta_{\text{road}}, \ \delta_{\text{road}} \in \{0,1\}.
\end{equation}
where $\delta_{\text{road}}$ is an indicator that transforms to $0$ when the EV is in an undrivable area. $\delta_{scal}$ is the scaling function that keeps the reward remains within the [-1,1] range.

\subsubsection{Efficiency Reward}
The efficiency reward consists of the overtake reward, the speed reward, and the exploration reward, which is expressed as follows:
\begin{equation}
    R_{\text{efficiency}} = R_{\text{speed}} + R_{\text{exploration}} + R_{\text{overtake}}.
\end{equation}

\noindent \textbf{Speed Reward} $R_{\text{speed}}$: The speed reward aims to attain the speed benefit for the EV. It is defined as:
\begin{equation}
    R_{\text{speed}} = \frac{v_{e,t}-v_{\text{thre}}}{||v_{\text{max}}-v_{\text{thre}}||},
\end{equation}
where $v_{e,t}$, $v_{\text{thre}}$ and $v_{\text{max}}$ is the velocity of the EV at time $t$, the threshold velocity above which the speed reward starts to grow and the maximum of velocity on the highway, respectively.

\noindent \textbf{Exploration Reward} $R_{\text{exploration}}$: The exploration reward encourages the EV to search for the space advantage through lane change. Specially, if the lane changes are too frequent that cause lateral oscillation, the reward will be negative to indicate discouragement. It is described as follows:
\begin{equation}
\begin{split}
&R_{\text{exploration}} = 
r_{\text{exp}} \ \ \text{if} \ \ a_t \ \ \text{in} \ A_{lc,t},\\
\mathcal{A}_{lc,t}=&\{a_t \in \mathcal{A}\setminus \text{LK} \ | \ (a_t=\text{LC} \rightarrow l_{e,t}>L_{lm}) \\ &\land (a_t=\text{RC} \rightarrow l_{e,t}<L_{rm})\},
\end{split}
\label{r_exp}
\end{equation}
where $r_{\text{exp}}$ is a positive constant. Notion $l_{e,t}$ is the lane the EV is on at time $t$, $L_{lm}$ and $L_{rm}$ are the index numbers of the left-most lane and right-most lane respectively.

\noindent \textbf{Overtake Reward} $R_{\text{overtake}}$: The EV will receive an overtake reward as depicted as follows:
\begin{equation}
    R_{\text{overtake}} = c_o*\frac{N_{a,t}-N_{a,t-1}}{N_{\text{env}}},c_o \in \{c_{\text{pos}},c_{\text{neg}}\},
\end{equation}
where $c_o$ is a constant, $N_{a,t}$ and $N_{a,t-1}$ is the number of vehicles that the EV is ahead of at timestep $t$ and $t-1$, respectively. $N_{\text{env}}$ is the estimated total vehicles based on the vehicle density.

\subsubsection{Safety Reward} To promote safe driving behaviors, our safety reward incorporates both collision and TTC considerations:
    \begin{equation}
        R_{\text{safety}} = R_{\text{collision}} + R_{ \text{ttc}}.
    \end{equation}

\noindent \textbf{Collision Reward} $R_{\text{collision}}$: A negative reward will be activated when a collision happens:
\begin{equation}
    R_{\text{collision}} = r_{c} \ \  \text{if collision is true else}\ \ 0,
\end{equation}    
where $r_c$ is a negative constant for collision penalty.

\noindent \textbf{TTC Reward} $R_{\text{ttc}}$: The longitudinal TTC reward and lateral TTC reward are included in the overall TTC reward, which also accounts for a distance penalty:
\begin{equation}
    R_{\text{ttc}} = R_{\text{lon.ttc}} + R_{ \text{lat.ttc}} + R_{\text{distance}},
\end{equation}    
where $R_{\text{lon.ttc}}$ account for the leader in the ego lane ($\text{el}$) and the leader in the target lane ($\text{tl}$) during lane change. The $R_{ \text{lat.ttc}}$ considers the neighboring vehicles in the lateral region of interest ($\text{nv}$). The last term in $R_{\text{distance}}$ represents a penalty based on the minimum absolute distance to any vehicle. Specifically, the rewards are defined as follows:
\begin{align}
R&_{\text{lon.ttc}} = -\frac{c_{\text{tl,lon}}}{t_{\text{tl}}} -\frac{c_{\text{el,lon}}}{t_{\text{el}}}, \quad  R_{\text{lat.ttc}} =-\frac{c_{\text{nv,lat}}}{t_{\text{nv}}}, \nonumber \\
R&_{\text{distance}} = -\frac{c_{\text{tl,lon}}}{d_{\text{tl}}} -\frac{c_{\text{el,lon}}}{d_{\text{el}}} -\frac{c_{\text{nv,lat}}}{d_{\text{nv}}} -\frac{c_{\text{min,dis}}}{d_{\text{min}}},
\end{align}
where $c_{\boxed{},\boxed{}}$ indicates constants for the rewards.

\subsection{LLM integrated Actor-Critic Architecture}
\label{subsec:actor-critic}
This subsection details our decision-making pipeline, which integrates a structured prompt with an actor head and a critic head that share a common LLM backbone.
\begin{figure}[t]
\centerline{\includegraphics[width=0.47\textwidth,height=0.47\textwidth]{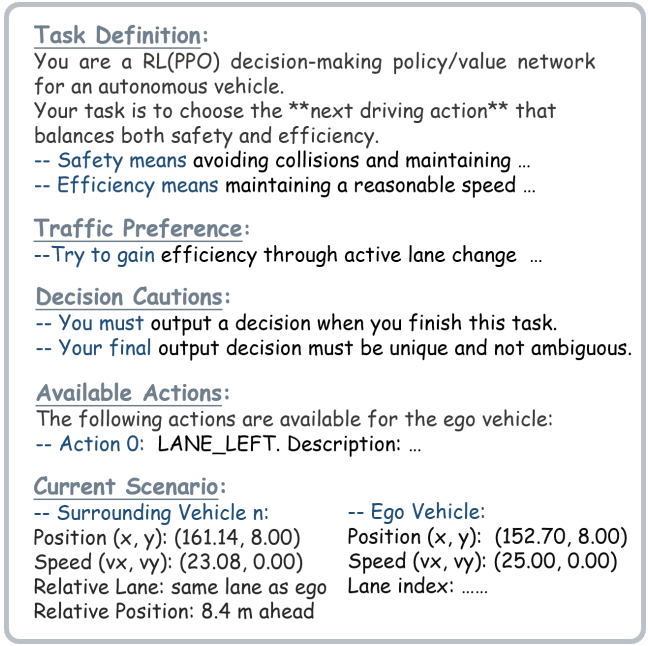}}
\caption{The format of the input prompt. It contains the task definition, traffic preference, available actions and current scenario, which helps the EV to understand the scenario and make environmentally informed decisions.}
\label{prompt}
\end{figure}
\subsubsection{Prompt Input Design}
The state of each time step, denoted as $\mathbf{S}_t=\{X_t,Y_t,V_{x,t},V_{y,t}\}$, consists of the states of the EV and a certain number of HDVs. All agents' positions with longitudinal and lateral velocities describes a scenario in one frame precisely.  According to the state given at each timestep, we construct an informative prompt for LLM to generate an effective decision. It consists of five main parts, including the \textit{Task Definition}, \textit{Traffic Preference}, \textit{Available Actions}, and \textit{Current Scenario}, as shown in Fig.~\ref{prompt}. The \textit{Task Definition} explains the role of the agent in this planning problem. The \textit{Traffic Preference} and \textit{Decision Cautions} specify driving inclination and provide reminders for making decisions in the mixed traffic scenarios. The \textit{Available Actions} limited the choice to a determined action selection from $\mathcal{A}=\{\textit{left change, lane keeping, right change}\}$, and the following text will use LC for \textit{left change}, LK for \textit{lane keeping}, and RC for \textit{right change} to keep the notation concise. Lastly, the \textit{Current Scenario} depicts essential information about the detailed situation of the EV. This structured prompt input provides the LLM with rich contextual information, allowing it to generate informed and instant decisions for the EV.

\subsubsection{Actor Head (Policy)} The tokenized text representation of the state is passed through the LLM backbone, which consists of frozen layers and trainable layers.

\noindent \textbf{Frozen layers}: The token embeddings derived from the tokenizer, denoted as $e_{\text{prompt}} = \operatorname{embedding}(\operatorname{prompt}(s))$, are first processed through the frozen layers of the LLM:
$h_{\text{frozen}} = \operatorname{LLM}_{1:N_f}(e_{\text{prompt}}; \theta_{\mathrm{frozen}})$.
Here, $N_f$ represents the number of frozen layers, and $\theta_{\mathrm{frozen}}$ denotes the parameters of these frozen network layers.

\noindent \textbf{Trainable layers}: The output from the frozen layers is then passed through the trainable layers for the driving task adaptation: $h_{\text{train}} = \mathrm{LLM}_{N_f+1:N_{\text{total}}}(h_{\text{frozen}}; \theta_{\mathrm{\text{train}}})$, the $N_{\text{total}}$ is the total number of LLM layers, $\theta_{\mathrm{train}}$ is the parameters of train layers' networks.

The retained knowledge and adaptation layers enable wise driving decision-making. Finally, the processed information is input to the multilayer perceptron (MLP) policy head, which outputs a probability distribution over the available actions:
\begin{equation}
\mathrm{logits} = \mathrm{MLP_{\text{policy}}}(h_{\text{train}}, \theta_P) \in \mathbb{R}^{|\mathcal{A}|},
\end{equation}
where $\theta_p$ is the parameter of policy network $\mathrm{MLP_{\text{policy}}}$, with
action generated as follows:
\begin{equation}
    \pi(a|s) = \frac{\exp(\mathrm{logits}(a))}{\sum_{a' \in \mathcal{A}} \exp(\mathrm{logits}(a'))}.
\end{equation}

\subsubsection{Critic Head (Value)}
The critic estimates the expected return of a given state, guiding the optimization of the policy network. To efficiently leverage the language representation, the critic shares the same backbone as the actor. After passing through the LLM processing layers, the output $h_{\text{train}}$ is passed to a separate MLP value head, which converts the extracted features into a scalar value estimate $V(s) = \mathrm{MLP_{\text{value}}}(h_{\text{train}}, \theta_V) \in \mathbb{R}^1,$ where $\theta_V$ is the parameter of value network $\mathrm{MLP_{\text{value}}}$.

According to the training procedure detailed in Section~\ref{sec:pre_RL_PPO}, the parameters $\theta_P$, $\theta_V$, and $\theta_{\text{train}}$ are updated based on the loss function presented in Eq.~\eqref{PPO_loss}.

\subsection{Safety-Critical Planner}
\label{subsec:safety-critical}
In this subsection, we will introduce the critical components of the planner, specifically detailing the constraints and objectives for the MPC framework.
\begin{figure}[t]
\centerline{\includegraphics[width=0.47\textwidth,height=0.10\textwidth]{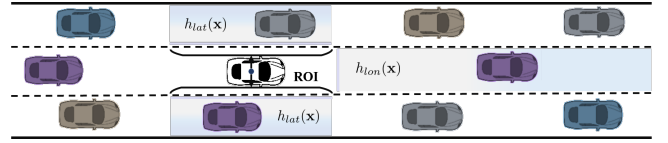}}
\caption{The Description of longitudinal DCBF and lateral DCBF within the range of ROI, which are critical components of safety-critical planner.}
\label{fig:lon_lat_dcbf}
\end{figure}

\subsubsection{Constraints}
The set of constraints comprises the system dynamics, longitudinal DCBF, lateral DCBF, and state limits.

\noindent \textbf{\textit{System Dynamics}}. For the system dynamics, we utilize a kinematic model, for which the mathematical expression  is described as follows: 
\begin{equation}
\label{eq:kinematic}
\begin{split}
v_{k+1} &= v_k + a_k \Delta t, \\
x_{k+1} &= x_k + v_k \cos(\theta_k) \Delta t, \\
y_{k+1} &= y_k + v_k \sin(\theta_k) \Delta t, \\
\psi_{k+1} &= \psi_k + \frac{v_t}{L} \tan(\delta_k) \Delta t,
\end{split}
\end{equation}
where the state vector $\left[x,y,v,\psi\right]^\top$ denotes the vehicle's longitudinal position, lateral position, velocity and yaw angle respectively. $L$ is the length of the vehicle. $\left[a,\delta\right]$ represents acceleration and steering angle. $\Delta t$ is the discerete time interval. Linearizing the kinematic model around the vehicle’s operating point like~\cite{sakai2018pythonrobotics} and using the Euler method, we can get a discretized model:
\begin{equation}
     \mathbf{x}_{k+1} = \bar{A}\mathbf{x}_k+ \bar{B}\mathbf{u}_k + \bar{C},
\end{equation}
where $\bar{A}$, $\bar{B}$ and $\bar{C}$ are system matrices.

\noindent \textbf{\textit{Longitudinal DCBFs}}. The longitudinal DCBF ensures that the vehicle maintains a safe distance from the vehicle ahead in the current lane, as shown in Fig.~\ref{fig:lon_lat_dcbf}. For this situation, a linear decoupled DCBF candidate is designed as follows:
\begin{equation}
    h_{lon}(\mathbf{x})=|x-x_i|- \alpha v - r_{lon},
    \label{Longitudinal Constraints}
\end{equation}
where $\alpha$ is a positive coefficient and $r_{lon}$ is a longitudinal safety distance. The longitudinal constraints can hence be constructed by Eq.~\eqref{pre:dcbf} with a slack mechanism as follows:
\begin{equation}
    \Delta h_{lon}(\mathbf{x}_k,\mathbf{u}_k) \geq -\gamma_{l} h_{lon}(\mathbf{x}_k) 
    +\epsilon_{lon}.
    \label{dcbflon}
\end{equation}
where the $\epsilon_{lon}$ is the longitudinal DCBF slack variable.

\noindent \textbf{\textit{Lateral DCBFs}}. In our design of lateral constraints, we incorporate the concept of the region of interest (ROI), as shown in Fig.~\ref{fig:lon_lat_dcbf}. The lateral constraint is triggered when the longitudinal gap between the ego vehicle and another vehicle in the adjacent lane becomes smaller than the defined ROI $|x-x_i| \leq r_{roi}$, where $r_{roi}$ is the longitudinal ROI value. In a manner similar to the longitudinal DCBF candidate, the lateral DCBF candidate $h_{lat}(\mathbf{x}) = |y-y_i| - r_{lat}$, the $r_{lat}$ is the lateral safety distance. The lateral constraints can then be formed the same way as:
\begin{equation}
    \Delta h_{lat}(\mathbf{x}_k,\mathbf{u}_k) \geq -\gamma_{l} h_{lat}(\mathbf{x}_k) + \epsilon_{lat},
\end{equation}
where the $\epsilon_{lat}$ is the lateral DCBF slack variable.

\noindent \textbf{\textit{State Limits}}. The state limits include constraints for velocity limits, control input limits and the limits of change of control inputs. We abbreviate it as $\mathbf{x}_{t+k+1|t} \in \mathcal{X}, \mathbf{u}_{t+k+1|t} \in \mathcal{U}, \dot{\mathbf{u}}_{t+k+1|t} \in \mathcal{U}_d$.

\subsubsection{Objectives}
The planner aims to minimize a cost function comprising several objectives, defined as follows:
\begin{equation}
\label{eq:cost}
\begin{split}
    J^*=&\sum_{k=0}^{N-1} ||\mathbf{u}_{t+k}||_Q + ||\dot{\mathbf{u}}_{t+k}||_P + \\
    &||\mathbf{x}_{t+k}-\mathbf{x}_{ref}||_R + p(\mathbf{x}_{t+N|t})_S + ||\epsilon||_{R_{\epsilon}}, \ \\    
\end{split}
\end{equation}
In this cost formulation, $N$ denotes the planning horizon. The cost function is composed of several weighted terms: $\sum_{k=0}^{N-1} ||\mathbf{u}_{t+k}||_Q$ penalizes the control effort, encouraging efficient actuator usage; $\sum_{k=0}^{N-1}||\dot{\mathbf{u}}_{t+k}||_P$ imposes a cost on the rate of change of the control inputs, thereby promoting smoother motion trajectories and $\sum_{k=0}^{N-1}||\mathbf{x}_{t+k}-\mathbf{x}_{ref}||_R$ quantifies the tracking error and efficiency, penalizing deviations from the reference state $\mathbf{x}_{ref}$. To enhance stability at the end of the horizon, the terminal cost $p(\mathbf{x}_{t+N|t})_S $ is formulated as a control Lyapunov function (CLF), specifically designed for the terminal yaw angle for stability. Additionally, the term $||\epsilon||_{R{\epsilon}}$ introduces a penalty on slack variables, which are incorporated to maintain the feasibility of the optimization problem under constrained scenarios. The weighting matrices $Q$, $P$, $R$, $S$, and $R_{\epsilon}$ are carefully tuned to balance the relative influence of these respective objectives.

We utilized the MPC framework as described in Eq.~\eqref{mpc} to form the optimization problem as follows:
\begin{equation}
\begin{split}
&\underset{\mathbf{u}_t:t+N-1|t,\mathbf{x}_t:t+N|t}{\arg\min} \sum_{k=0}^{N-1} ||\mathbf{u}_{t+k}||_Q + ||\dot{\mathbf{u}}_{t+k}||_P + \\
&\quad ||\mathbf{x}_{t+k}-\mathbf{x}_{ref}||_R + p(\mathbf{x}_{t+N|t})_S + ||\epsilon||_{R_{\epsilon}}, \ \text{s.t.} \\
&\mathbf{x}_{t+k|t} = \bar{A}_{t+k|t} \mathbf{x}_{t+k|t} + \bar{B}_{t+k|t} \mathbf{u}_{t+k|t} + \bar{C}_{t+k|t}, \\
&\Delta h_{lon}(\mathbf{x}_{t+k|t}, \mathbf{u}_{t+k|t}) + \gamma_h h_{lon}(\mathbf{x}_{t+k|t}) \geq \epsilon_{lon}, \\
&\Delta h_{lat}(\mathbf{x}_{t+k|t}, \mathbf{u}_{t+k|t}) + \gamma_l h_{lat}(\mathbf{x}_{t+k|t}) \geq \epsilon_{lat}, \\
&\mathbf{x}_{t+k+1|t} \in \mathcal{X}, \mathbf{u}_{t+k+1|t} \in \mathcal{U}, \dot{\mathbf{u}}_{t+k+1|t} \in \mathcal{U}_d, \\
&\mathbf{x}_{t|t} = \mathbf{x}_{t}, \, k = 0, \dots, N-1.
\end{split}
\label{eq:optimization}
\end{equation}
The optimal solution is $\mathbf{u_t^{*}}=[\mathbf{u}_{t,0}^*, ..., \mathbf{u}_{t,N-1}^*]$, the first element $\mathbf{u}_{t,0}^*$ is applied to Eq.~\eqref{eq:kinematic} to get the new state $\mathbf{x}_{t+1}$.

\section{Experiment}
\label{sec:experiment}
In this section, we begin by outlining the implementation details in Section~\ref{sec:Implementation Details}, followed by a discussion of the metrics and baselines in Section~\ref{sec:Metrics and Baselines}. Finally, the detailed performance evaluation is presented in Section~\ref{sec:Performance Evaluation}.
\begin{table}[t]
\centering
\captionsetup{
  justification=centering, 
  labelsep=space, 
  textfont=sc, 
  labelfont=sc, 
  format=plain 
}
\caption{Parameters in LA-RL}
\label{tab:parameters}
\renewcommand{\arraystretch}{1.05} 

\begin{tabularx}{\linewidth}{@{}lXc@{}}
\toprule
Symbol & Description & Value \\ \midrule
$v_{\text{thre}}$ & Threshold velocity &  20 \si{m/s} \\
$v_{\text{max}}$ & Maximum of velocity on the highway & 30 \si{m/s} \\
$r_{\text{exp}}$ & Positive constant reward for exploration& 0.4  \\
$c_{\text{pos}}$ & Constant for positive overtake reward &  5 \\
$c_{\text{neg}}$ & Constant for negative overtake reward & 0.5 \\
$r_c$ & Constant for negative collision reward & -2  \\
$c_{\text{tl,lon}}$ & Constant for TTC reward (target lane)  & 0.2 \\
$c_{\text{el,lon}}$ & Constant for TTC reward (ego lane) & 0.2 \\
$c_{\text{nv,lat}}$ & Constant for TTC reward (neighbor lane) & 0.2 \\
$c_{\text{min,dis}}$ & Constant for TTC reward (minimum distance) & 0.2 \\
$Q$ & Weight for control inputs & $[0.05 0.05]_{\text{diag}}$ \\
$P$ & Weight for jerk &  $[0.2 0.2]_{\text{diag}}$ \\
$R$ & Weight for state difference & $[0 8.0 0.1 0]_{\text{diag}}$\\
$S$ & Weight for terminal cost & 5.0 \\
$R_{\epsilon}$ & Weight for slack cost & $[10,500]_{\text{diag}}$ \\
$\gamma_h$ & Coefficient for longitudinal DCBF & 0.8 \\
$\gamma_l$ & Coefficient for lateral DCBF & 0.8 \\
$N$ & Predictive horizon & 10 \\
$L$ & The length of the vehicle & 5 \\
$\Delta t$ & Discerete time interval & 0.2 \\
$\gamma$ & Discount factor & 0.8 \\
$\lambda$ & Coefficient for GAE & 0.95 \\
$\epsilon$ & Surrogate clipping coefficient & 0.2 \\
$c_2$ &  Coefficient of entropy loss & 0.0 \\
$c_1$ & Coefficient of value function & 0.5 \\
\bottomrule
\end{tabularx}
\end{table}

    





\begin{table*}[t]
\centering
\caption{Comparison of Multiple Metrics under different vehicle densities.}
\footnotesize
\setlength{\tabcolsep}{7.5pt}
\footnotesize
\begin{tabular}{>{\centering\arraybackslash}m{1.4cm}|>{\centering\arraybackslash}m{1cm}|cccccccc}
\hline
\rowcolor{gray!10} 
Method & Density & TTC S. $\uparrow$& S.R.  $\uparrow$& Avg. Pro.  $\uparrow$&   Max. Pro.  $\uparrow$&Avg. Vel $\uparrow$&  Avg. Jerk $\downarrow$&  Avg. Acc. $\downarrow$& Avg. L.C. $\uparrow$\\
\hline

\multirow{3}{*}{\makecell[c]{ DQN \\  (MLP)}} 
& Low &\textbf{2.1594} &  67.00 & 1037.26& \textbf{1506.0}9 &\textbf{22.2311} & 4.3606 & 1.0515  &   \textbf{2 (1.78)}\\
& Medium &1.9985  & 24.00 & 439.45 &1210.00&20.0859* &  9.0048& 1.3644 &\textbf{4 (3.72)}\\
& High&  1.7059& 7.00 &164.93&1219.95 & 19.5474* & 18.4072 & 2.5803   &  1 (0.91)\\
\hline

\multirow{3}{*}{\makecell[c]{ DQN \\   (CNN)}} 
& Low & 0.9553&  4.00 & 557.61&\textbf{1674.25}& 26.0904* &11.1705  & 1.7784 &   \textbf{3 (2.55)}\\
& Medium &  2.3651 & 0.00 & 124.34 &406.23&21.0853* &20.5517  & 2.5290 &  1 (1.33)\\
& High&  1.9526&  0.00 &  53.79*&187.41&  19.7551* & 38.2798 & 3.4946 & 1 (0.77)\\
\hline

\multirow{3}{*}{\makecell[c]{ PPO \\ (MLP)}}
&  Low & \textbf{2.9342}  & 60.00& 928.87&\textbf{1340.60}&21.6233 &2.6733 & 0.3825 &  0 (0.44)\\
& Medium& 1.6860&  12.00& 282.85 &1254.29& 20.0629*& 11.6740 & 1.6518 &  0 (0.20)\\
& High & 1.7991  &5.00 &109.69 &1205.39&19.1381* & 23.9956 &3.4715  & 0 (0.04)\\
\hline

\multirow{3}{*}{\makecell[c]{ PPO \\ (Attention)}} 
& Low & 33.3315**  &94.00** & 1150.23**&1180.51& 19.9961**&  \textbf{0.0614}**&  \textbf{0.1180}**&   0 (0.06)**\\
& Medium& 3.5296& 22.00 & 199.85 & 1183.00&  9.5966*&  2.2938&  0.9917 & 0 (0.26)\\
& High  & 2.7981 &  2.00 &39.43 &1180.51  &9.1367*  & 4.2211 & 1.9092&  0 (0.12)\\

\hline
\multirow{3}{*}{\makecell[c]{SO-DM \\ System}} 
& Low& 7.1861 & 32.00 & 583.71 &  1401.71& 24.2435* & 1.4525 & 1.3746 & 0 (0.58)\\
& Medium &2.2958  &  3.00&  119.08&1148.66  & 23.8302* & 4.1397& 2.2277 &  0 (0.08) \\
& High & 1.5855  & 0.00 &42.71  &  202.91& 24.2242*& 3.5187 &1.8692 &0 (0.00)  \\
\hline

\multirow{3}{*}{\makecell[c]{DRB-FSM \\ System }} 
& Low& 6.1684  & 29.00 &  547.83&1460.02  &24.0808*  & 1.4529 &1.4490&0 (0.57)  \\
& Medium&  2.2034&  10.00 & 204.93 & 1241.78 &23.5628*  & 3.0203 &  1.7982 & 0 (0.14)\\
& High & 1.8101 &  2.00 & 60.40 & 1164.54 & 24.3262*&2.5215 &1.4360  &  0 (0.01) \\
\hline





 \cellcolor{cyan!5}& \cellcolor{cyan!5}Low& \cellcolor{cyan!5}\textbf{2.4845}& \cellcolor{cyan!5}\textbf{100.00} & \cellcolor{cyan!5}\textbf{1322.27} & \cellcolor{cyan!5}\textbf{1564.34}  & \cellcolor{cyan!5}\textbf{22.4403} & \cellcolor{cyan!5}\textbf{0.3258}  &  \cellcolor{cyan!5}\textbf{0.7521} & \cellcolor{cyan!5}\textbf{4 (3.87)}\\
 \cellcolor{cyan!5}LA-RL & \cellcolor{cyan!5}Medium& \cellcolor{cyan!5}\textbf{3.2273}  & \cellcolor{cyan!5}\textbf{98.00} &  \cellcolor{cyan!5}\textbf{1113.34} & \cellcolor{cyan!5}\textbf{1296.76} & \cellcolor{cyan!5}\textbf{19.1618}  &\cellcolor{cyan!5}\textbf{0.0142} &  \cellcolor{cyan!5}\textbf{0.9474} &  \cellcolor{cyan!5}\textbf{4 (3.80)}\\
 \cellcolor{cyan!5}& \cellcolor{cyan!5}High & \cellcolor{cyan!5}\textbf{2.9808} & \cellcolor{cyan!5}\textbf{84.00}& \cellcolor{cyan!5}\textbf{956.00} & \cellcolor{cyan!5}\textbf{1360.19} & \cellcolor{cyan!5}\textbf{17.8157} & \cellcolor{cyan!5}\textbf{0.7000} & \cellcolor{cyan!5}\textbf{1.3944} &\cellcolor{cyan!5}\textbf{3 (2.65)} \\
\hline


\hline
\end{tabular}
\begin{tablenotes}    
        \footnotesize              
        \item[1] Note: \textbf{Bold} indicates good performance, * represents potentially less meaningful data, and ** indicates strongly correlated data.

\end{tablenotes} 

\label{tab:results}
\end{table*}

\begin{figure*}[t]
\caption{A comparison of multiple methods based on success steps, evaluated under different settings.}
\centerline{\includegraphics[width=\textwidth,height=0.16\textwidth]{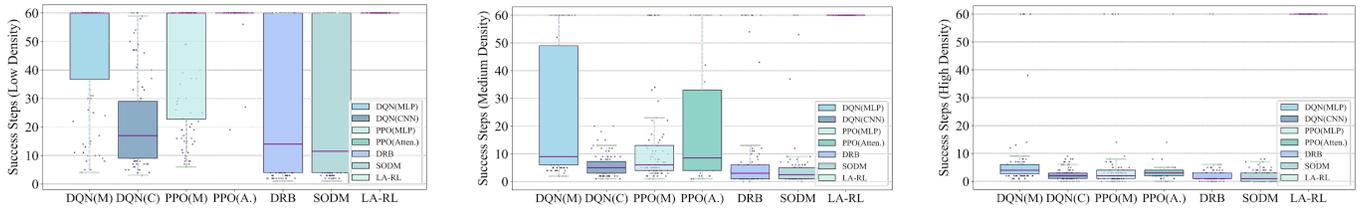}}
\label{boxplot}
\end{figure*}

\subsection{Implementation Details}
\label{sec:Implementation Details}
In our experiments, we implemented and trained the algorithms using Stable-Baselines3~\cite{stable-baselines3} in conjunction with Gymnasium~\cite{towers2024gymnasium} environments, facilitating effective experimentation. To evaluate LA-RL's planning performance, we employ the \textit{highway-v0} scenario from highway-env \cite{highway-env}. The training batch size is set to 32, and the learning rate is configured to 5 $\times$ 10$^{-4}$. We utilize an NVIDIA GeForce RTX 4090 GPU for both training and inference. The model is trained for a total of 40,000 timesteps. Additionally, the LLM integrated into the LA-RL framework is SmolLM2~\cite{allal2025smollm2}, with the SmolLM2-135M-Instruct variant selected for implementation with its final layer unfrozen during
training. The parameters of LA-RL is listed in Table.~\ref{tab:parameters}.

\begin{figure*}[t]
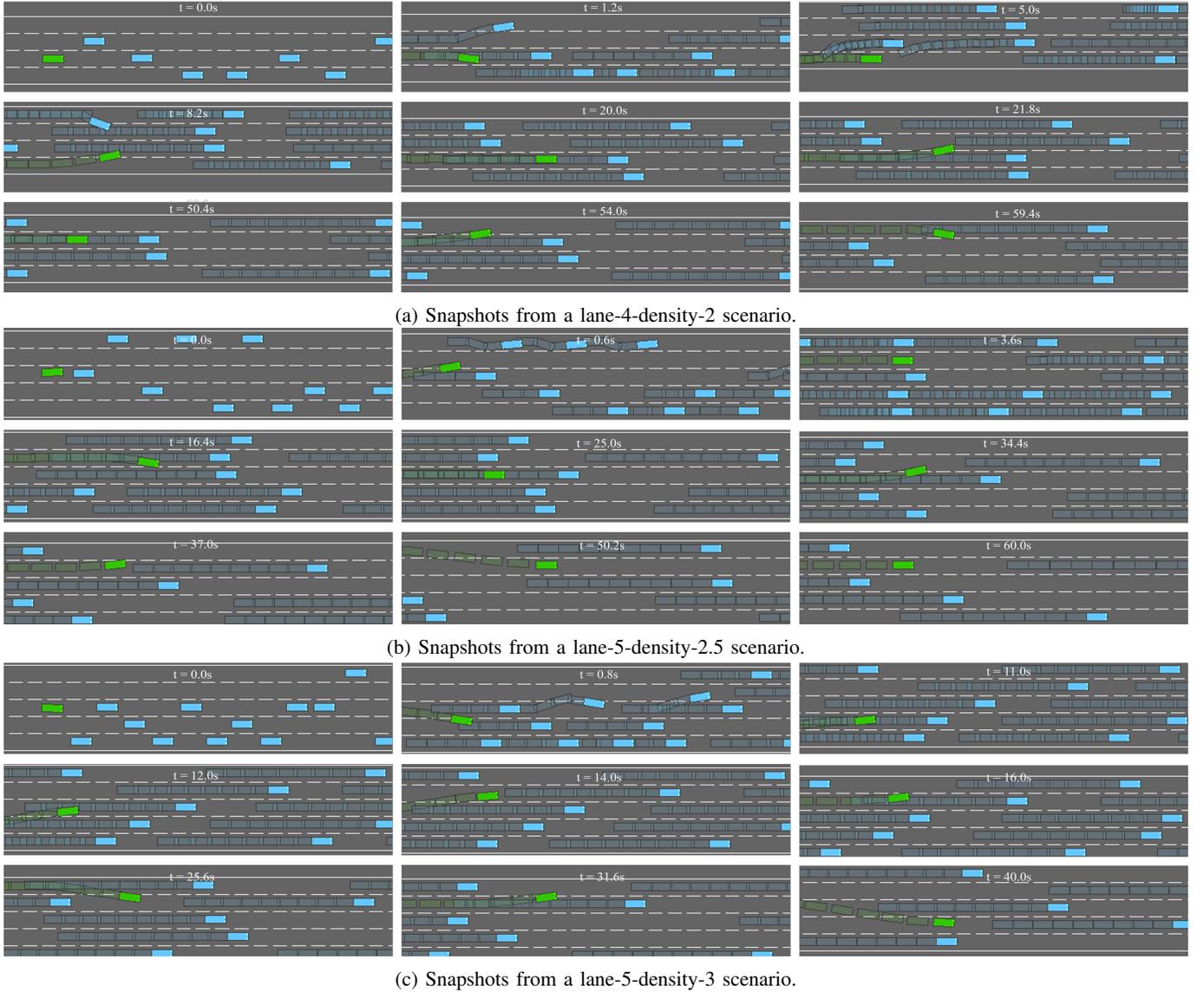

    \centering
    \begin{subfigure}[b]{\textwidth}
        \includegraphics[width=1\textwidth,height=0.25\textwidth]{lane4d2.pdf}
        \caption{Snapshots from a lane-4-density-2 scenario.}
        \label{fig:density1}
    \end{subfigure}
    \hfill
    \begin{subfigure}[b]{\textwidth}
        \includegraphics[width=1\textwidth,height=0.25\textwidth]{lane5d25.pdf}
        \caption{Snapshots from a lane-5-density-2.5 scenario.}
        \label{fig:density2}
    \end{subfigure}
    \begin{subfigure}[b]{\textwidth}
        \includegraphics[width=1\textwidth,height=0.25\textwidth]{lane5d3.pdf}
        \caption{Snapshots from a lane-5-density-3 scenario.}
        \label{fig:density3}
    \end{subfigure}
    
    \caption{Snapshots of scenarios from various configuration settings.}
    \label{fig:snapshots}
\end{figure*}

\subsection{Metrics and Baselines}
\label{sec:Metrics and Baselines}


\subsubsection{Baselines} We compare LA-RL with numerous baselines, providing a rich set of comparative results. The following is a list of all the baseline methods used in the comparison:

\noindent \textbf{PPO} \cite{schulman2017proximal,highway-env}: Two types of Proximal Policy Optimization will be compared: one using an MLP-based feature extractor and the other utilizing a Transformer with attention mechanisms.

\noindent \textbf{DQN} \cite{highway-env} Two variants from \cite{highway-env} are selected as baselines, each employing different architectures: one based on an MLP and the other on a CNN.

\noindent \textbf{SO-DM} \cite{shu2023safety} The SO-DM model encourages autonomous vehicles to prioritize increasing speed while keeping safety in mind.

\noindent \textbf{DRB-FSM} \cite{chen2020hierarchical} The FSM-based decision-making approach for DBR determines suitable driving behaviors through state transitions, effectively coordinating vehicle motion and improving overall driving flexibility.

\noindent \textbf{DiLu}  \cite{wen2024dilu} DiLu is a knowledge-driven autonomous driving framework that uses reasoning, reflection, and memory to improve decision-making.

\noindent \textbf{GRAD} \cite{xi2022graph} GRAD is an RL method that utilizes a space-time graph to represent vehicle interactions and future trajectories for autonomous driving tasks.

\noindent \textbf{Hybrid-Driving} \cite{wang2025hybrid} Hybrid-Driving integrates LLMs, knowledge graphs, and driving rules to improve decision-making in autonomous driving and address LLM hallucination issues.

\noindent \textbf{ADRD} \cite{zeng2025adrd} ADRD is a framework that combines large language models with rule-based decision systems for interpretable autonomous driving.

\subsubsection{Metrics} We compare the methods mentioned above in two kinds of settings. The first setting consists of multiple \textit{three-lane scenarios} and we evaluated them over a $60\si{s}$ duration across 300 parallel tracks. These tracks were distinctly divided into three vehicle density settings, with 100 tracks allocated to each setting. The second setting includes \textit{challenging scenarios} tailored for various configurations, specifically lane-4-density-2.0, lane-5-density-2.5, lane-5-density-3.0 for 30$\si{s}$, the same experimental setup for state-of-the-art works: DiLu, GRAD Hybrid-Driving, ADRD and we will compare their strengths as highlighted in the respective papers with LA-RL. In the \textit{three-lane scenario} configuration, models for the baseline methods and LA-RL were individually trained under each density condition. In \textit{challenging scenarios}, each model was trained utilizing the lane-4-density-2.0 setting, mirroring the exact setup employed for training GRAD in the DiLu.
This comprehensive evaluation enables us to assess LA-RL's performance across various comparisons. The evaluation metrics are summarized as follows:


\noindent\textbf{TTC Score:} We use the time-to-collision scores (TTC S.) as the key safety indicator, with higher values indicating better performance, while excessively high TTC scores may indicate an overly conservative strategy.

\noindent\textbf{Progress:} Final progress is a key measure of the vehicle's ability to move forward, providing insight into its effectiveness in navigating various scenarios.

\noindent\textbf{Efficiency:} Efficiency is crucial, and we use average speed as the metric. However, it only holds value when the success rate is high, as a low success rate typically indicates collisions, making the average speed less meaningful in evaluation.

\noindent\textbf{Comfort:} To measure comfort, we evaluate the vehicle's dynamics through the absolute average acceleration and jerk across all tracks, providing insights into ride smoothness and stability.

\noindent\textbf{Lane Change Intention:} Lane change intention evaluates how well policies choose the right moments for lane changes to optimize space and speed.

\noindent\textbf{Success Situation:} The success rate (SR) is the most vital metric, where each track is deemed successful unless a collision or unsolvable issue occurs. Success steps (SS) represent the number of successful steps within a given duration.

\noindent\textbf{Control Efficiency:} Control efficiency is the average inference time per decision (seconds per command) as defined in~\cite{zeng2025adrd}.

\subsection{Performance Evaluation}
\label{sec:Performance Evaluation}
\begin{table*}[t]
\caption{Ablation Study Results of LA-RL.}
\centering
\footnotesize
\setlength{\tabcolsep}{10pt}
\footnotesize
\begin{tabular}{>{\centering\arraybackslash}m{1.5cm}|>{\centering\arraybackslash}m{1.0cm}|cccccccc}
\hline
\rowcolor{gray!10} 
 Method& Setting & TTC S. & S.R.  & Avg. Pro.  &   Max. Pro.  &Avg. Vel &  Avg. Jerk &  Avg. Acc. & Avg. L.C. \\
\hline

\multirow{3}{*}{\makecell[c]{LA-RL}} 
& L4-D2.0 &9.7317 &97.50 &579.31 & 737.02& 20.6317& 0.9548& 1.5324& 3 (3.33)\\
& L5-D2.5 &9.1045 &92.50 &551.46& 850.08&20.6030 & 1.0979&  1.7471& 9 (9.20)\\
& L5-D3.0& 8.9178&80.00 &497.40 & 793.52& 20.5757&1.3905 & 2.6226& 8 (8.28)\\
\hline

\multirow{3}{*}{\makecell[c]{LA-RL \\Simple Reward}} 
& L4-D2.0 &11.7144 &92.50 &539.21 &666.91 & 20.2396& 0.9634&1.3686 &2 (2.025) \\
& L5-D2.5 &8.9938 &85.00 & 519.88& 745.17& 20.6846& 1.3202& 2.1978& 12 (12.58)\\
& L5-D3.0& 10.7614&80.00 &484.78 &763.76& 20.0835&2.0069 & 2.3358& 12 (12.45)\\
\hline



\multirow{3}{*}{\makecell[c]{LA-RL \\Less Slack}} 
& L4-D2.0 &7.9319 &90.00 & 566.45& 812.06&21.6888 &1.0609 &1.8025 &3 (3.15) \\
& L5-D2.5 & 8.0689& 87.50& 555.51& 838.13& 21.4174& 1.1694& 2.1948& 6 (5.93)\\
& L5-D3.0& 7.7781& 77.50&500.55 &794.23 & 21.2418&1.0910 & 2.6327& 7 (6.85)\\
\hline

\end{tabular}
\label{tab:results_ablation}
\end{table*}

\begin{table}[h]
\centering
\setlength{\tabcolsep}{14pt} 
\caption{Lane Change Intention (Low Density)}
\begin{tabular}{@{}lccc@{}}
\toprule
Method & Left L.C.& Right L.C.& Max. L.C.\\ \midrule
DQN (MLP) & 1 (0.74) & 1 (1.04) & \textbf{11}\\
DQN (CNN) & 1 (1.20)&  1 (1.35)& \textbf{12}\\
PPO (MLP) &0 (0.07)  & 0 (0.37) & 2\\
SO-DM & 0 (0.34) &0 (0.24)  &2 \\
DRB-FSM & 0 (0.34) &  0 (0.23)& 3\\
LA-RL  &\textbf{2 (2.04)} &  \textbf{2 (1.83)}& \textbf{11}\\
\bottomrule
\end{tabular}
\label{tab:Lane Change Intention}
\end{table}

\subsubsection{Three-lane Scenarios}
We compare the baselines in group one first. As shown in Table~\ref{tab:results}, LA-RL achieved a remarkably high SR in extremely challenging tasks, significantly outperforming all other baselines with improvements of  100$\%$, 98.00$\%$, 84.00$\%$. This SR is the most important indicator among all metrics, highlighting the practicality and superiority of LA-RL. Furthermore, as demonstrated in Fig.~\ref{boxplot}, LA-RL attains a notably high number of SS. The 60$s$ planning horizon presents a considerable challenge, evidenced by the relatively low SR achieved across all baselines, particularly under medium and high density conditions ranging from 0.00$\%$ to 24.00$\%$. Crucially, in the low-density regime, PPO(Attention) stands out as the only comparable baseline with a SR of 94$\%$. 

The high SR of PPO(Attention) in the low-density scenarios is strongly correlated with its other metrics marked with two asterisks. Specifically, its TTC score is extremely high at 33.3315, while the average number of lane changes is nearly zero. Maintaining low average speed and low average jerk, this suggests that PPO(Attention)'s strategy is primarily passive car-following, lacking an active and intelligent lane-changing policy to gain speed and space advantages. This conservative behavior is also reflected in its low maximum progress.

The data marked with an asterisk is considered less meaningful due to the extremely low SR achieved by the corresponding methods. Taking DQN (CNN) as a prime example, its SR across the low, medium, and high density settings is exceptionally low. This poor performance is coupled with a very high average speed and high average jerk. This indicates that the DQN (CNN) agent's behavior is often characterized by extremely rapid velocity changes. In most instances, the agent maintains a high speed and collides with the preceding vehicle, or it crashes during an aggressive lane-changing maneuver, an outcome we strongly want to avoid in the model's decision-making.

Both the SO-DM System and DRB-System exhibit low SR, a deficiency primarily stemming from their underlying planners lacking a slack mechanism. Consequently, in increasingly dynamic and dense scenarios, the planner often fails to slow down in time or the optimization solver becomes infeasible, ultimately leading to mission failure. This inherent limitation also accounts for why these systems maintain a relatively high average speed despite their poor SR. Focusing on the low-density scenario, the TTC scores for the SO-DM System and DRB-System reach 7.1861 and 6.1684, respectively. These data are notably high and further suggests that during their limited successful steps, these approaches tend towards conservative car-following behavior rather than executing proactive maneuvers.

Comparing their performance, LA-RL significantly outperforms these baselines in terms of both average progress and maximum progress, exceeding their results by a margin of approximately 100 to 200 meters. Additionally, LA-RL achieves a much higher average lane change count, reaching 3$\sim$4 times, also shown in Table~\ref{tab:Lane Change Intention}, a maximum number of lane changes compared to some relatively good lane-changing ability. It demonstrates that, while maintaining high SR, LA-RL makes wise lane change decisions to gain speed and space benefits. Furthermore, under active exploration, LA-RL achieves a reasonable TTC score, complemented by lower jerk and average acceleration values.


\subsubsection{Challenging Scenarios}

\begin{figure}[t]
\centerline{\includegraphics[width=0.53\textwidth,height=0.32\textwidth]{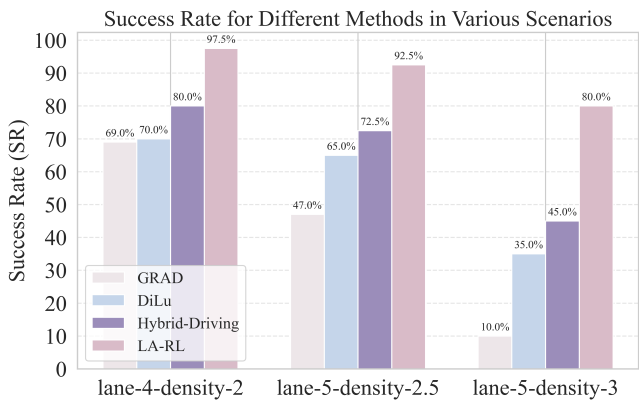}}
\caption{A success rate comparison of LA-RL, Hybrid-Driving, DiLu, and GRAD across various scenario configurations.}
\label{sr_group_2}
\end{figure}

In comparison with DiLu, GRAD, and Hybrid-Driving, Fig.~\ref{sr_group_2} demonstrates that Hybrid-Driving significantly outperforms the existing state-of-the-art approaches in terms of SR. In three different scenario settings, Hybrid-Driving achieves success rates of 80$\%$, 72.5$\%$, and 45$\%$, respectively. Specifically, GRAD exhibits the weakest performance, with its SR plummeting to a mere 10.0$\%$ in the highest density setting. Similarly, DiLu's SR peaks at 70.0$\%$ but degrades substantially with increasing traffic density, reaching only 35.0$\%$ in the most challenging scenario. Notably, LA-RL surpasses Hybrid-Driving by approximately 20$\%$ across all settings, with a remarkable 97.5$\%$ SR in the lane-4-density-2.0 setting, 92.5 $\%$ SR in the lane-5-density-2.5 setting and 80.0$\%$ in the lane-5-density-3.0 setting.

\begin{table}[t]
\centering
\footnotesize

\setlength{\tabcolsep}{8pt}
\caption{Comparison of Avg. Dri. Time and Ctr. Effi.}
\footnotesize
\begin{tabular}{>{\centering\arraybackslash}m{1.23cm}|>{\centering\arraybackslash}m{0.95cm}|ccc}
\hline
Setting & Method & Avg. Dri. Time (s) & Ctr. Eff. (s/c.)\\
\hline
\multirow{3}{*}{\makecell[c]{Lane-4 \\ density-2.0}}
& DiLu&23.00&  14.33 \\
&  ADRD&25.15  & $\mathbf{< 10^{-6}}$\\
& LA-RL&  \textbf{29.33}& \textbf{0.0279}\\
\hline

\multirow{3}{*}{\makecell[c]{Lane-5 \\ density-2.5}} 
& DiLu & 16.00& 12.42\\
& ADRD & 16.75 &$\mathbf{< 10^{-6}}$\\
& LA-RL&  \textbf{27.88}&   \textbf{0.0288}\\
\hline

\multirow{3}{*}{\makecell[c]{Lane-5 \\ density-3.0}}
&  DiLu & 10.10& 12.67\\
& ADRD& 13.55& $\mathbf{< 10^{-6}}$\\
& LA-RL & \textbf{24.38} &\textbf{0.0283}\\

\hline
\end{tabular}
\label{tab:control_efficiency}
\end{table}

As shown in the snapshots of Fig.~\ref{fig:density1}, the EV accelerates after securing a larger spatial margin. At $t=8.2s$, the EV executes a lane change to follow a faster leader, successfully completing an overtaking maneuver. Subsequently, as the separation distance increases between its new, speedier leader and the vehicle in the adjacent lane, the EV performs another lane change at $t=21.8s$ to follow an even faster leader. Following this, the EV executes a final lane change to complete a subsequent overtaking, thereby maximizing both its spatial and speed gains. In the snapshots of Fig.~\ref{fig:density2}, the EV starts out following another vehicle with an exceptionally tight gap. The EV quickly responds by braking and moving to an adjacent lane. The policy then continues to make effective and intelligent lane changes to secure better gains in speed and spatial margin. The scenario depicted in Fig.~\ref{fig:density3} features extremely dense and heavily congested traffic flow. Despite these challenging conditions, the LA-RL policy demonstrates exceptional robustness. The EV is still able to execute judicious lane change decisions at the right moments, successfully achieving spatial and speed benefits while navigating the high-density environment.

Table~\ref{tab:control_efficiency} compares the metrics in ADRD, showing that LA-RL significantly outperforms DiLu and ADRD in average driving time with values of 29.33$\si{s}$, 27.88$\si{s}$, and 24.38$\si{s}$, respectively. Additionally, ADRD has the lowest average inference time, while DiLu's inference time exceeds 10$\si{s}$. LA-RL, with an inference time of around 0.028$\si{s}$, along with ADRD, supports real-time decision-making.

\subsubsection{Ablation Studies}
The ablation study, summarized in Table \ref{tab:results_ablation}, compares the performance of the LA-RL framework against two variants: LA-RL with simple reward and LA-RL with less slack. The LA-RL with simple reward variant utilizes the original reward function as defined in the \textit{highway-env} environment~\cite{highway-env}. The LA-RL with less slack variant modifies the planner by reducing the weight of the slack variable penalty within the optimization objective.

The performance of LA-RL with simple reward is degraded compared to the standard LA-RL model, as evidenced by a reduction in its SR. Specifically, the SR drops to 92.50$\%$ under the lane-4-density-2 scenario and to 85.00$\%$ under the lane-5-density-2.5 scenario. The decrease is more pronounced in average progress, and the maximum progress lags behind that of LA-RL by approximately 100 meters. This reduction in performance is simultaneously accompanied by an increase in average jerk. This suggests that while the SR is lower, the policy also achieves less spatial benefit and provides reduced ride comfort.
The LA-RL policy implemented with reduced slack shows a marginal decrease in its SR. Despite this, its progress metrics are comparable to the original LA-RL, as is the average speed. The reduced penalty for slack allows the planner greater freedom for acceleration. This faster, more aggressive movement leads to trade-offs, including a slight increase in the average jerk and the reduction in the SR. Consequently, these findings indicate that the slack mechanism is quite crucial for ensuring policy safety. The results further demonstrate that the full LA-RL method generally achieves the best balance between safety and efficiency. 

\section{Conclusion}
\label{sec:conclusion}
This paper presented LA-RL, a novel reinforcement learning framework that integrates Large Language Models for semantic reasoning with a safety-critical MPC-DCBF planner for autonomous highway driving. This synthesis enables LA-RL to achieve superior adaptability and robustness in its driving performance. This synthesis of high-level environmental understanding and formal safety guarantees enables LA-RL to achieve a superior balance between proactive efficiency and robust safety. Extensive experiments conducted in diverse highway scenarios demonstrate that the proposed LA-RL framework significantly outperforms existing state-of-the-art methods across key driving metrics, including safety, efficiency, and comfort. It effectively navigates the the trade-off between exploiting driving performance and adhering to stringent safety constraints. This work establishes a promising direction for building reliable autonomous agents that can reason and act effectively in complex environments. Future work will extend this paradigm by modeling the adversarial and cooperative interactions, enabling our agent to understand the environment and anticipate other drivers' intentions to develop strategic, socially-compliant driving behaviors that anticipate the intentions of other agents.



%
\bibliographystyle{IEEEtran}
\bibliography{ref.bib}

\vspace{-1cm}
\begin{IEEEbiography}
[{\includegraphics[width=1in,height=1.25in,clip,keepaspectratio]{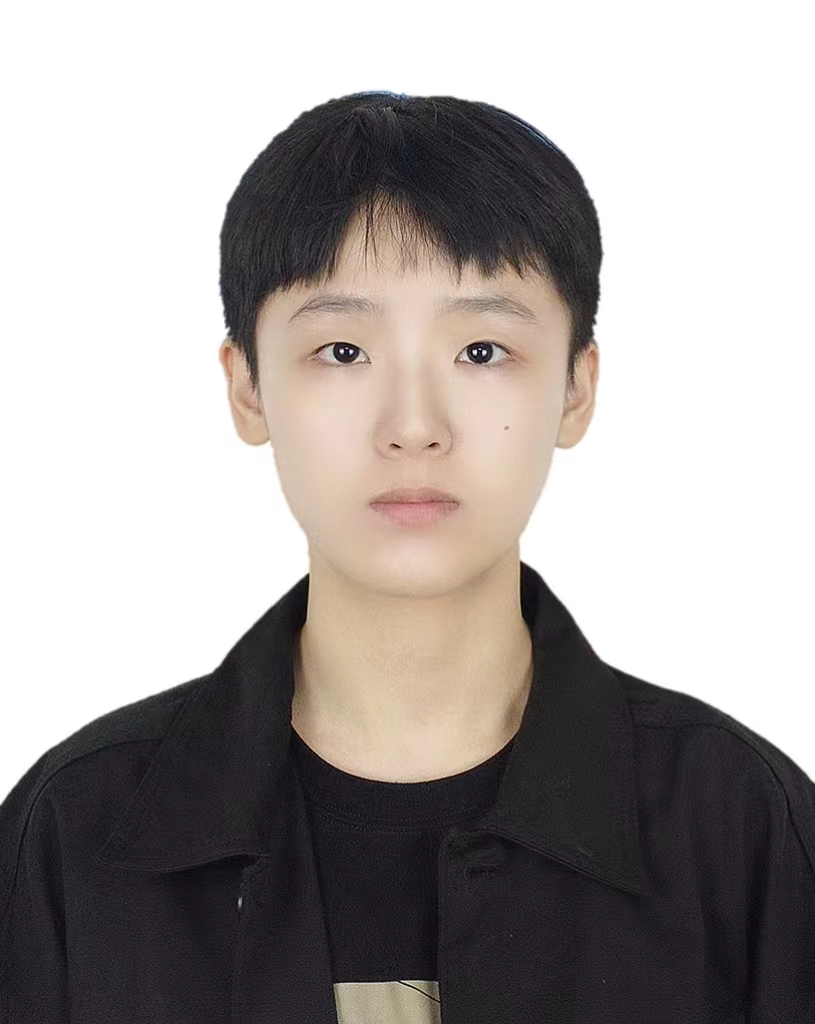}}]{Yiming Shu}
received the B.Eng. degree in Automotive Engineering from the Harbin Institute of Technology (Weihai), Weihai, China, in 2022, the  M.Phil. degree in 2025 from the University of Hong Kong (HKU). She is currently pursuing her Ph.D. in the Department of Data and Systems Engineering at the University of Hong Kong. Her research interests include motion planning and decision-making of autonomous vehicles (AVs).
\end{IEEEbiography}
\vspace{-1cm}

\begin{IEEEbiography}
[{\includegraphics[width=1in,height=1.25in,clip,keepaspectratio]{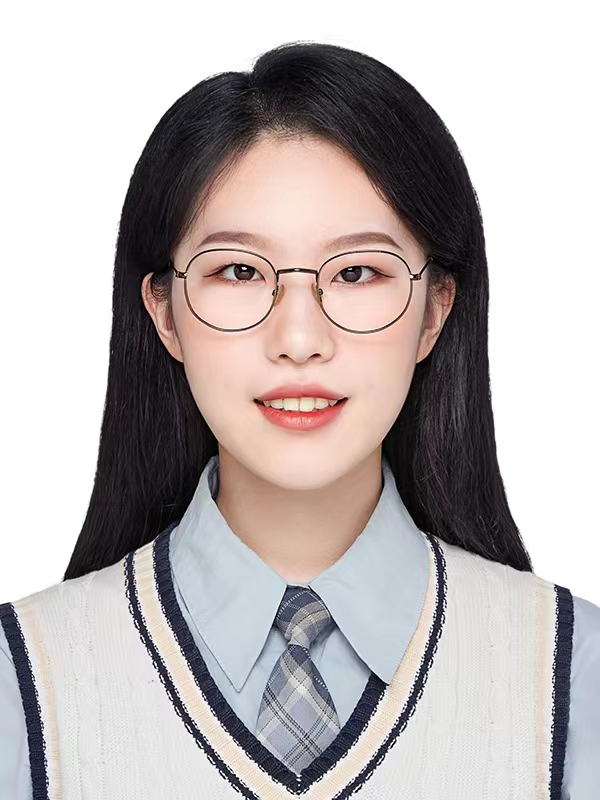}}]{Jiahui Xu}
received her Bachelor's degree in Vehicle Engineering from Beijing Institute of Technology, China in 2022, and the Master's degree in 2025. She is currently pursuing her Ph.D. in the Department of Data and Systems Engineering at The University of Hong Kong. Her research interests include trajectory prediction, decision-making, and the safety of autonomous driving.
\end{IEEEbiography}
\vspace{-1cm}

\begin{IEEEbiography}[{\includegraphics[width=1in,height=1.25in,clip,keepaspectratio]{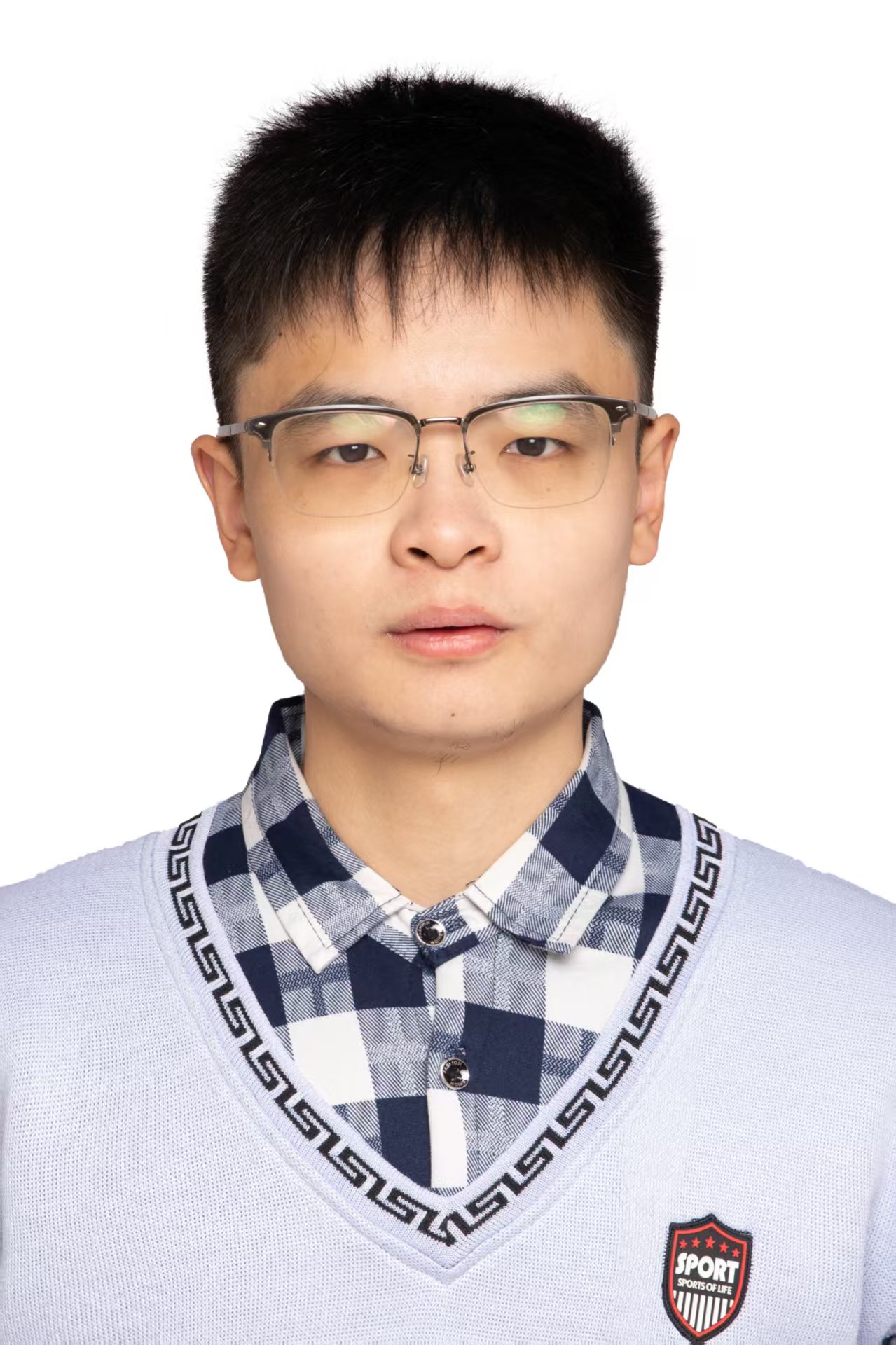}}]{Jiwei Tang} obtained his Bachelor's degree in Marine Engineering and Technology from Zhejiang University, Hangzhou, China in 2022, and Master's degree in Electrical and Computer Engineering from National University of Singapore, Singapore in 2024. He is currently pursuing his Ph.D. in the Department of Data and Systems Engineering at The University of Hong Kong. His research interests include nonlinear system control, constrained control, deep reinforcement learning, and autonomous systems.
\end{IEEEbiography}
\vspace{-1cm}

\begin{IEEEbiography}[{\includegraphics[width=1in,height=1.25in,clip,keepaspectratio]{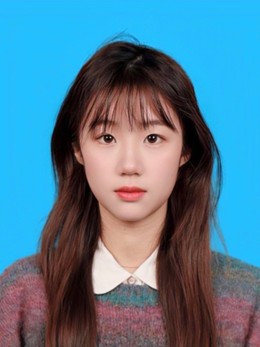}}]{Ruiyang Gao} received the Bachelor degree from Beijing University of Posts and Telecommunications, Beijing, China in 2024 and M.S. degree in Artificial Intelligence with the College of Computing and Data Science, Nanyang Technological University, Singapore, Singapore in 2025. She is currently working toward the Ph.D. degree with the Department of Data and Systems Engineering, The University of Hong Kong, Hong Kong SAR, China. Her research interests include intelligent cockpit, human-machine interaction, and affective computing.
\end{IEEEbiography}
\vspace{-1cm}

\begin{IEEEbiography}[{\includegraphics[width=1in,height=1.25in,clip,keepaspectratio]{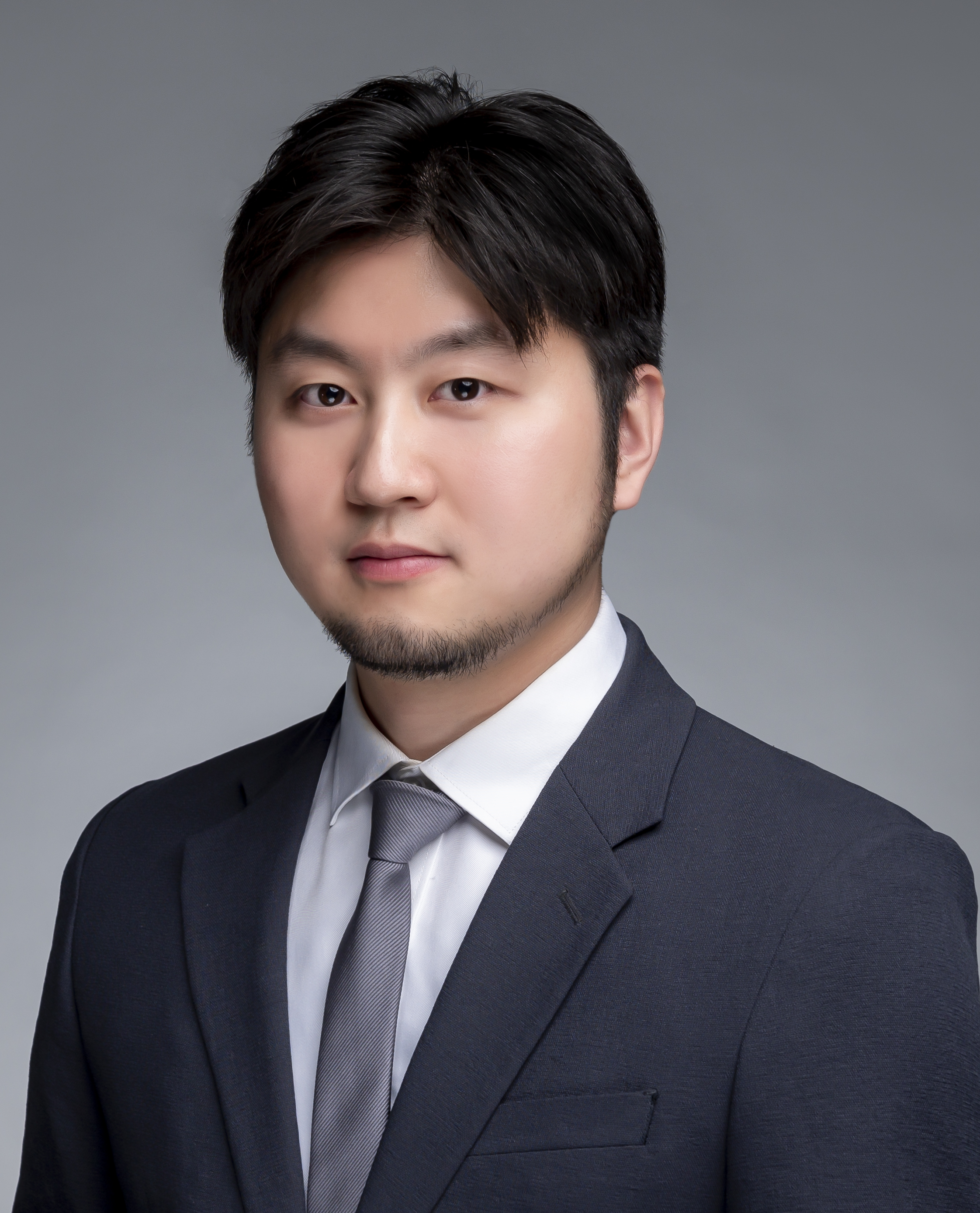}}]{Chen Sun}
received the Ph.D. degree in Mechanical
\& Mechatronics Engineering from University of
Waterloo, ON, Canada in 2022, M.A.Sc degree in
Electrical \& Computer Engineering from University
of Toronto, ON, Canada in 2017 and B.Eng. degree
in automation from the University of Electronic
Science and Technology of China, Chengdu, China,
in 2014. He is currently an Assistant Professor
with the Department of Data and Systems Engineering, University of Hong Kong. His research
interests include field robotics, safe and trustworthy
autonomous driving and in general human-CPS autonomy.
\end{IEEEbiography}

%








\end{document}